\documentclass[12pt,leqno]{amsart}
\usepackage{amsmath,amsthm,amssymb}

\def\init{\setcounter{equation}{0}}
\newtheorem{theorem}{Theorem}[section]

\newcommand{\e}{{\varepsilon}}

\def\init{\setcounter{equation}{0}}

\newcommand{\R}{\mathbb{R}}

\newcommand{\rw}{\rightarrow}

\usepackage{graphicx,color}
\newlength{\figboxwidth}             
\newcommand{\makefig}[3]{
        \begin{figure}[htb]
        \refstepcounter{figure}
        \label{#2}
        \begin{center}
                #3~\\
                \smallskip
                Figure \thefigure.  #1
        \end{center}
        \medskip
        \end{figure}
}

\title{
Nonstationary analogue  black holes
}

\author{
Gregory Eskin, Department of Mathematics, UCLA,\\
Los Angeles, CA 90095-1555, USA
}

\begin{document}
\maketitle
\begin{abstract}
We study  the existence of analogue nonstationary spherically 
symmetric black holes. 
The prime example is the acoustic model (cf. [V], [U]).  We consider 
also a more general  class
of  metrics that could be useful  in other physical models of analogue black and white 
holes.
 We give examples of the appearance of black holes
and of disappearance of white holes.  We also discuss the relation
between the apparent  and the event horizons for the case of analogue black holes.
In the end we study the inverse  problem  of determination  of black or  white holes  by
boundary measurements for the spherically  symmetric nonstationary metrics. 
\end{abstract}

\section{Introduction}
\init
Let
\begin{equation}                               \label{eq:1.1}
\sum_{j,k=0}^n g_{j,k}(x_0,x)dx_jdx_k
\end{equation}
be
a Lorentzian metric with the signature $(1,-1,...,-1)$,  where\linebreak 
$x=(x_1,...,x_n)\in\R^n,\ \ x_0\in \R$  is 
the time variable.  We assume that 
$g_{jk}(x_0,x)\in C^\infty(\R^n\times\R),\ \ g_{jk}(x_0,x)
=\eta_{jk}+O\big(\frac{1}{|x|}\big)$ uniformly in $x_0,$
\linebreak  
$[\eta_{jk}]_{j,k=0}^n$  is 
the Minkowski tensor. 
We also assume 
that $g_{jk}(x_0,x)-g_{j,k}(\pm\infty,x)$  decay fast
uniformly in $x$  when $x_0\rw\pm\infty$.

Consider the wave equation
\begin{equation}                                   \label{eq:1.2}
\Box_g u(x_0,x)\overset{\text{def}}{=} \sum_{j,k=0}^n
\frac{1}{\sqrt{(-1)^ng(x_0,x)}}
\frac{\partial}{\partial x_j}
\Big(\sqrt{(-1)^ng}\ g^{jk}(x_0,x)\frac{\partial u}{\partial x_k}\Big)=0,
\end{equation}
where $[g^{jk}]_{j,k=0}^n=\big([g_{jk}]_{j,k=0}^n\big)^{-1},\
 g(x_0,x)=\det[g_{jk}]_{j,k=0}^n$.

In the case when the metric $[g_{jk}(x_0,x)]_{j,k=0}^n$  is a solution of
the Einstein's equations of  general relativity, the solutions $u(x_0,x)$  of the equation (\ref{eq:1.2})
describe linear gravitational waves in the background of the metric (\ref{eq:1.1}).

One of the striking phenomenons of  general relativity is the appearance of black holes  
and white holes.

In general relativity  the domain $D\subset\R^n\times\R$  is called a black hole if 
no signal (disturbance)  inside $D$  can escape to the spatial infinity.

In this paper we shall consider some classes of nonstationary black and white holes that  arise
in the study  of analogue  space-times.  
Our definitions are similar 
but different from the definition of black and white holes in general relativity 
(cf. [W], [FN]).

{\bf Definition 1.1}
Let $D$  be a domain in $\R^n\times\R$  such that 
the domains $D_t=D\cap\{x_0=t\}\subset\R^n$  are uniformly bounded on $\R$.
Suppose $\partial D$  is a characteristic surface  for the equation (\ref{eq:1.2}).

We say  that $D$ is a  black hole if no signal (disturbance) 
 inside $D$  can reach  the exterior  of $D$.

{\bf Definition 1.2}
The domain $D\subset\R^n\times\R$  is called an
outer  black hole  
  if it is a  black hole and any signal (disturbance)  outside $D$
reaches spatial infinity when $x_0\rw +\infty$.

The boundary  of a  black hole (outer black hole) is called  a
black hole horizon (outer black hole horizon).
The Definition 1.2 
is similar to the definition of a black hole in general  relativity.

Analogously,  $D$  is a  white hole if no signal (disturbance)  outside $D$  can
penetrate into the interior  of $D$,  and $D$ is an outer white hole  if it
is  a  white  hole and any signal (disturbance) outside  $D$  can reach  
the spatial infinity when $x_0\rw -\infty$.
  We introduced the notion of an outer black  hole because 
 sometimes there exists more then one
  black hole.  
Then only the largest   black hole that contains all other will
be the outer black hole.  Moreover,  the  outer black holes
can be determined  by the boundary measurements  (cf. \S5).

The equation (\ref{eq:1.2}) describes also linear waves in a moving 
medium such as the propagation of light in moving dielectric (cf. [G], [LP],
[NVV]) or acoustic waves in a moving fluid (cf. [V1], [U], [BLV]).  The black  
holes for such equations are called optical  or acoustic black holes.
They are also called analogue or artificial black holes.  
There are other interesting physical  models of analogue spacetimes   as surface waves
models, Bose-Einstein condensates,  gravity waves  and others  (see excellent  papers 
[V1],  [V2],  where many physical  examples  are considered). 
The study of analogue black holes
is much simpler than the study of the black holes in the general relativity and
physicists  expect that it will help them to understand better the black holes
of the general relativity.  They also expect to study the analogue  black 
holes experimentally. 
Another road leading to the study of analogue black holes is 
the attempt to answer the following question (cf. [E2], [E4]):
Can one determine the coefficients of the wave equation by the boundary
measurements?
Such problems  are called the inverse hyperbolic problems.  The presence
of black hole will make the determination of
   the coefficients  inside 
the black hole impossible. 
\enlargethispage{\baselineskip}

\vbox{
Now we shall describe the content of the present paper. 
In \S 2 we study  the event and apparent horizons and the relations between them.
 In \S 3 we shall
refine the results of [E1] on the existence of  black and white holes
for stationary metrics in  two space dimensions. The novelty here
is the consideration of metric with singularity at some point.
Although the emphasis in the paper  is on  the spherically  symmetric case we
consider  in \S3  the nonspherically symmetric metrics too, since the proofs in the
stationary  case are almost the same.  
In \S 4 we study the existence of nonstationary
 black and white holes depending on $x_0$  and $|x|$ only.
Note  that  there are only  few studies of  nonstationary analogue black holes,  mostly
in the case  of one space dimension (cf. [BLSV]). 
We consider separately the case of acoustic black or white holes 
and the case of  general  nonstationary spherically symmetric metrics.
We give also an example of black holes appearing at some time $x_0=t$
and white  holes disappearing at some $x_0=t$.
Finally,  in \S5  we consider the inverse problem  of the determination of a
nonstationary spherically symmetric black or
a white hole knowing  the time-dependent  Dirichlet-to-Neumann operator on the boundary.
Inverse problems for stationary  metrics in any dimensions  were considered in [E2].
\section{Event horizon and apparent horizon}
\init

In this section we describe black and white holes analytically, 
 extending the proof
of Theorem 3.1 of [E1]  to the case of nonstationary metrics. 
We assume that
\begin{equation}                                  \label{eq:2.1}
g^{00}(x_0,x)>0,\ \ \ \forall (x_0,x).
\end{equation}                 
Let $S(x_0,x)=0$  be a smooth  surface in $\R^n\times\R,$
closed in $\R^n,\ \forall x_0\in \R$  and the gradient
$(S_{x_1},...,S_{x_n})$  be not zero when $S(x_0,x)=0$.  We shall
choose outward direction of the gradient.
Suppose $S(x_0,x)=0$  is a characteristic surface for the equation (\ref{eq:1.2}),
i.e. 
\begin{equation}                                     \label{eq:2.2}
\sum_{j,k=0}^n g^{jk}(x_0,x)S_{x_j}(x_0,x)S_{x_k}(x_0,x)=0\ \ \mbox{when\ }\ 
S(x_0,x)=0.
\end{equation}
Equation (\ref{eq:2.2})  has two roots with respect to $S_{x_0}$:
\begin{gather}                                \label{eq:2.3}
S_{x_0}^\pm=\big(-\sum_{j=1}^n g^{j0}(x_0,x)S_{x_j}\pm \sqrt{Q(x_0,x)}\big)/{g^{00}(x_0,x)},
\ \ \mbox{where}\ \ \ \ \ \ \   
\\ 
Q(x_0,x)=\Big(\sum_{j=1}^ng^{j0}(x_0,x)S_{x_j}\Big)^2-
g^{00}(x_0,x)\sum_{j,k=1}^ng^{jk}(x_0,x)S_{x_j}S_{x_k}.
\end{gather}
}
Since (\ref{eq:1.2}) is strictly hyperbolic we have that
$
Q(x_0,x)>0 \ \ \ \forall(x_0,x).
$
Therefore
$
S_{x_0}^-(x_0,x)<S_{x_0}^+(x_0,x).
$
\begin{theorem}                                   \label{theo:2.1}
Let $d_{x_0}$ be the diameter of $S(x_0,x)=0$  in $\R^n$  for
fixed $x_0$.  Suppose that $d_{x_0}$ is bounded on $(-\infty,+\infty)$.  Suppose 
$S(x_0,x)$  is a smooth characteristic  surface,
$-\infty<x_0<+\infty$.
  Then $S(x_0,x)=0$  is a boundary of a  black hole 
if $S_{x_0}(x_0,x)$ is the smaller  root in (\ref{eq:2.3}), 
 i.e. $S_{x_0}(x_0,x)=S_{x_0}^-(x_0,x),$
and $S(x_0,x)=0$  is a boundary of a  white hole if $S_{x_0}(x_0,x)$  is the larger root
in (\ref{eq:2.3}), i.e. $S_{x_0}(x_0,x)=S_{x_0}^+(x_0,x)$.
\end{theorem}
{\bf Proof:} Let $S(y_0,y)=0$, and let $S_{x_0}=S_{x_0}^+$  be the larger root in (\ref{eq:2.3}).
Denote by $K_+(y_0,y)$  the forward time-like half-cone in $\R^n\times \R$  consisting of 
$(\dot x_0,\dot x_1,...,\dot x_n)\in \R^n\times\R$
such that 
\begin{equation}                                   \label{eq:2.5}
\sum_{j,k=0}^n g_{jk}(y_0,y)\dot x_j\dot x_k>0,\ \ \dot x_0>0.
\end{equation}
Let $K^+(y_0,y)$ be the half-cone of the dual cone 
\begin{equation}                                   \label{eq:2.6}
\sum_{j,k=0}^n g^{jk}(y_0,y)\xi_j\xi_k>0,\ \ \ (\xi_0,...,\xi_n)\in \R^n\times\R,
\end{equation}
that contains $(1,0,...,0)$.  Note that 
\begin{equation}                                   \label{eq:2.7}
\sum_{k=0}^n\dot x_k\xi_k>0
\end{equation}
for all $(\dot x_0,...,\dot x_n)\in K_+(y_0,y)$  and all 
$(\xi_0,\xi_1,..,\xi_n)\in K^+(y_0,y)$.
Vector $(S_{x_0}^++\e,S_{x_1}^+,...,S_{x_n}^+)\in K^+(y_0,y)$ for any $\e>0$ 
because
\begin{multline}                          
\nonumber
g^{00}(y_0,y)(S_{x_0}^++\e)^2+2\sum_{j=1}^ng^{j0}(y_0,y)(S_{x_0}^++\e)S_{x_j}^+
+\sum_{j,k=1}^n g^{jk}(y_0,y)S_{x_j}^+S_{x_k}^+
\\
=g^{00}(y_0,y)\e^2+2\e\Big(\sum_{j=0}^ng^{j0}S_{x_j}^+\Big).
\end{multline}
We use that $S^+$  satisfies (\ref{eq:2.2}).  It follows from
(\ref{eq:2.3}) that 
$
\sum_{j=0}^ng^{j0}S_{x_j}^+=+\sqrt Q>0.
$  
Thus
$
g^{00}\e^2+2\e\sqrt Q>0.
$
Therefore
\begin{multline}                                           \label{eq:2.8}
g^{00}(S_{x_0}^++\e)^2+2\sum_{j=1}^ng^{j0}(S_{x_0}^++\e)S_{x_j}^+
\\
+\sum_{j,k=1}^ng^{jk}(y_0,y)S_{x_j}^+S_{x_k}^+>0,\ \ \forall \e >0.
\end{multline}
Taking $\e>0$  large we get that the vector 
$$
(S_{x_0}^++\e,S_{x_1}^+,...,S_{x_n}^+)
=\e\Big(1+\frac{S_{x_0}^+}{\e},\frac{S_{x_1}^+}{\e},...,\frac{S_{x_n}^+}{\e}\Big)
$$ 
can
be deformed continuously to the vector
$\e(1,0,...,0)\in K^+(y_0,y)$.  Therefore 
$(S_{x_0}^++\e,S_{x_1}^+,...,S_{x_n}^+)$ belongs to the half-cone  $K^+(y_0,y)$.
It follows from (\ref{eq:2.7}) that
\begin{equation}                                  \label{eq:2.9}
(S_{x_0}^++\e)\dot x_0+\sum_{k=1}^nS_{x_j}^+\dot x_k>0
\end{equation}
for all $(\dot x_n,...,\dot x_n)\in K_+(y_0,y),\ \forall\e>0$.
Passing to the limit when $\e\rw 0$  we get
\begin{equation}                               \label{eq:2.10}
\sum_{j=0}^nS_{x_j}^+\dot x_j\geq 0
\end{equation}
for all $(\dot x_1,...,\dot x_n)\in \overline K_+(y_0,y)$.  Since
 $(S_{x_0}^+,...,S_{x_n}^+)$  is an outward normal the inequality (\ref{eq:2.10})
shows that all forward time-like rays starting at $(y_0,y)$ are pointed inside 
the exterior of $S^+(x_0,x)=0$.  This proves that $S^+(x_0,x)=0$ is a boundary 
of a  white hole.

Consider now  the case when $S(x_0,x)=S^-(x_0,x_0)$,  i.e.
$S_{x_0}^-$ is the smaller 
root in (\ref{eq:2.3}).  For $\e>0$  we have using  (\ref{eq:2.3})
that
$(-S_{x_0}^-+\e,-S_{x_1}^-,...,-S_{x_n}^-)\in K^+(y)$
and as in (\ref{eq:2.9}),  (\ref{eq:2.10})  we get
that
\begin{equation}                              \label{eq:2.11}
\sum_{j=0}^nS_{x_j}^-\dot x_j\leq 0
\end{equation}
for all $(\dot x_0,...,\dot x_n)\in K_+(y)$.
Since $(S_{x_0}^-,...,S_{x_n}^-)$
is an outward normal to $S^-(y_0,y)=0$, 
 we have that any forward time-like ray starting at $(y_0,y)$
is pointed inside $S^-(x_0,x)=0$.  Therefore  $S^-(x_0,x)=0$  is a boundary
of the  black hole.
\qed

Black or white hole  horizons 
are the notions global in time $x_0$.  The introduction of the notion 
of an apparent horizon is a way to get an information  about the event
horizon (black or white hole) at a given time.  

Consider a nonstationary metric in $\R^n\times\R$. Fix $x_0=t$.  Let
$S(x)=0$ be a closed smooth surface in $\R^n\times\{x_0=t\}$.
We assume  that $S_x(x)\neq 0$  when  $S(x)=0$  and that $S_x(x)$ 
 is an outward normal to $S(x)=0$.  

Consider  the system  of null-bicharacteristics (cf. [E1])
with initial data
$
 x_0(0)=t,  x_j(0)=y_j, 1\leq j\leq n,  S(y)=0,
 \xi_j(0)=S_{x_j}(y), 1\leq j\leq n.
$
 There are two such null-bicharacteristics with
\begin{equation}                                     \label{eq:2.12}
\xi_0^\pm(0)=\frac{1}{g^{00}(t,y)}
\ \Big(
-\sum_{j=1}^ng^{j0}(t,y)S_{x_j}(y)\pm\sqrt{Q(t,y)}\Big),
\end{equation}
where $Q(t,y)$  is the same as in (\ref{eq:2.4}).
Note that the strict hyperbolicity  implies that
$Q(t,y)>0$.  Thus if
\begin{equation}                                  \label{eq:2.13}
\sum_{j,k=1}^ng^{jk}(t,y)S_{x_j}(y)S_{x_k}(y)\geq 0
\end{equation}
then $Q>0$ implies  
$
\sum_{j=1}^ng^{j0}(t,y)S_{x_j}(y)\neq 0.
$
We assume in the case of black holes (cf. [E1])
that 
\begin{equation}                                  \label{eq:2.14}
\sum_{j=1}^ng^{j0}(t,y)S_{x_j}(y)< 0 \ \ \mbox{when}\ \ S(y)=0.
\end{equation}
In the case of white hole  we assume   that (\ref{eq:2.13})
holds  and
\begin{equation}                                  \label{eq:2.15}
\sum_{j=1}^ng^{j0}(t,y)S_{x_j}(y)> 0 \ \ \mbox{when}\ \ S(y)=0.
\end{equation}
Let  $K_+(y)$  and  $K^+(y)$  be the same  as in (\ref{eq:2.5}),  (\ref{eq:2.6}).
If (\ref{eq:2.13})  and  (\ref{eq:2.14}) hold then
\begin{equation}                                \label{eq:2.16}
\xi_0^+(0)>0,\ \ \ \xi^-(0)\geq 0.
\end{equation}
Note that $\xi^-(0)=0$  if  
\begin{equation}                                  \label{eq:2.17}
\sum_{j,k=1}^ng^{jk}(t,y)S_{x_j}(y)S_{x_k}(y)= 0.
\end{equation}
We generalize the proof of Theorem \ref{theo:2.1} for the case
when (\ref{eq:2.17})  may not hold.  

As in (\ref{eq:2.8}), (\ref{eq:2.9}),  (\ref{eq:2.10})
  we get  that $\overline{K_+}(y)\subset \Pi^+$,
where $\Pi^+$  is the half-space
\begin{equation}                                \label{eq:2.18}
\xi_0^+\dot x_0+\sum \dot x_kS_{x_k}(y)\geq 0.
\end{equation}
Analogously to (\ref{eq:2.11})  we get
that
  $\overline{K_+}(y)\subset \Pi^-$ where $\Pi^-$  is the half-space
\begin{equation}                                    \label{eq:2.19}
\xi_0^-\dot x_0+\sum_{k=1}^n\dot x_kS_{x_k}(y)\leq 0.
\end{equation}
\vbox{
Thus $\overline{K_+}(y)\subset\Pi^+\cap\Pi^-$.
Denote 
by $D\subset \R^n$  the domain with boundary  $\{S(x)=0\}$.
If $\xi_0^+(0)>0,\ \xi_0^-(0)\geq 0$  for all $y$  such that 
$S(y)=0$
the domain $D\times\R$  is a trapped (no escape)  region for the stationary
metric
\begin{equation}                                      \label{eq:2.20}
\sum_{j,k=0}^ng_{jk}(t,x)dx_jdx_k,
\end{equation}  
where $t$  is fixed  (``frozen").
When 
$\xi_0^-(0)=0\ \ D\times\R$  is the outmost trapped  region,  i.e.
a  black hole  for the  metric (\ref{eq:2.20}),  where $t$  is  fixed.

Analogously,  when
(\ref{eq:2.13})
and 
(\ref{eq:2.15})  hold,  we have $\xi_0^-(y)<0,\linebreak \xi_0^+(y)\leq 0,\  S(y)=0$.
As in (\ref{eq:2.18}),  (\ref{eq:2.19})  we get  that
\begin{equation}                                   \label{eq:2.21}
\overline{K_+}(y)\subset \Pi_1^-\cap\Pi_1^+,
\end{equation}
where  $\Pi_1^-$  consists  of 
$(\dot x_0,...,\dot x_n), \dot x_0>0$,  satisfying \\ 
$
\xi_0^-(y)\dot x_0 +\sum_{k=1}^n\dot x_k S_{x_k}(y) \leq 0
$
and \\
$
\Pi_1^+=\{(\dot x_0,...,\dot x_n): \xi_0^+\dot x_0+
\sum_{k=1}^n\dot x_kS_{x_k}(y)\geq 
0,\ x_0>0\}.
$ \\
Therefore for the metric  (\ref{eq:2.20}), $t$  is fixed,  we have that any 
disturbance   
in the exterior  of $D\times\R$  can not reach the interior  of $D\times \R$,
i.e.  $D\times \R$  is an antitrapped  region.  In the case when
$\xi_0^+(y)=0\ \ D\times\R$  is the outmost antitrapped  region,  i.e. 
$S\times\R$  is a  white hole horizon  for the metric (\ref{eq:2.20}). 
Therefore  the apparent horizon is the black 
or white hole horizon  for the metric (\ref{eq:2.20}) when $x_0=t$  is fixed.
Note that $D\times \R$  is not necessary  a trapped region for
the non-stationary metric.
Examine the relation between the apparent horizon 
 and the event 
  horizon.
Let $S(x_0,x)=0$  be a smooth surface in $\R^{n+1}$  such that  
$S(x_0,x)=0$  is a smooth  closed
surface  in $\R^n$  for  each $x_0\in \R$.  Assume
that
$\xi_0^+(0)>0,\ \xi_0^-(0)=0$   (cf.  (\ref{eq:2.12}))  hold for each  $x_0$
when $S(x_0,y)=0$.  Then
$\{S(t,y)=0\}\times\R$  is the apparent  horizon  for the metric  
(\ref{eq:2.20}) when $x_0=t$  is fixed.
The surface $S(x_0,x)=0$ in $\R^{n+1}$  is called  the dynamic  horizon.
In \S 3 we shall study the 
stationary acoustic  metrics  and in \S 4 the nonstationary
acoustic metrics with  $A(x_0)\leq  A_0<0$  (cf. (\ref{eq:3.1}) and Theorem \ref{theo:4.1}).
Let  $r=r^+(x_0)$  be  the black  hole  horizon (cf. Theorem \ref{theo:4.1}).\linebreak
  The dynamic horizon
in this case is $r=|A(x_0)|$,  i.e
$\{r=|A(t)|\}\times\R$  is apparent  horizon  for each  $t$.  If 
$|A(x_0)|$  is increasing  when $x_0$  is increasing  we have that  the 
dynamic horizon is inside the black hole
with the boundary $r=r^+(x_0)$.
If  $|A(x_0)|$  is decreasing when $x_0$ is increasing  then the black  hole  with  
the boundary  $r=r^+(x_0)$  is  inside the domain  bounded by the dynamic horizon.

\section{Stationary black holes in two space dimensions}
\init
\subsection{Metric singularity}
In this section we consider the case of $n=2$  and
the Lorentz metric tensor $[g_{jk}(x)]_{j,k=0}^2$  independent  of 
the
time
 variable $x_0$.}

Let $S$  be the ergosphere (cf [E1]),  i.e the smooth closed curve $\Delta(x)=0$ 
where $\Delta(x)=g^{11}(x)g^{22}(x)-(g^{12}(x))^2$.

Suppose that $S$ is the boundary of  simply connected domain $\Omega$ and 
suppose $O=(0,0)\in \Omega$.  We assume that $[g_{jk}(x)]_{j,k=0}^n$  are $C^\infty$  
in $\Omega\setminus O,\ \Delta(x)<0$  in $\Omega\setminus O$  and
$[g_{jk}(x)]_{j,k=0}^2$ has a singularity at $O$.
Note that the Schwarzschield metric (see, for example, [W]) has a singularity at 
the origin.

Another example of metric having a singularity is the acoustic metric (cf. [V1])  when
\begin{equation}                                       \label{eq:3.1}
g^{00}=1,\  g^{j0}=g^{0j}=v^j,\ 1\leq j\leq 2,
\ \ \
g^{jk}=-\delta_{ij}+v^iv^j,\ 1\leq j,k\leq 2,
\end{equation}
where
$v=(v^1,v^2)$   is the velocity of the flow,
\begin{equation}                               
\nonumber
v=\frac{A}{r}\hat r + \frac{B}{r}\hat\theta,\ \ r=|x|,\ \ \hat r=\frac{x}{|x|},
\ \ \hat\theta=\frac{(-x_2,x_1)}{|x|},
\end{equation}
$A$  and  $B$  are constants.
Note that many metrics such as Kerr metric (cf. [W]), Gordon metric (cf. [G], [LP]) have
the form (\ref{eq:3.1}).

In case of general stationary metric in $\R^2\times\R$ we assume that when
$|x|<\e,\ \e$ is small,  the metric tensor $[g^{jk}]_{j,k=0}^2$  has the following form
\begin{equation}                                  \label{eq:3.2}
g^{jk}(x)=g_1^{jk}(x)+g_2^{jk}(x),\ \ |x|<\e,
\end{equation}
where $g_1^{jk}$  is similar to an acoustic metric,
\begin{align}                                  \label{eq:3.3}
&g_1^{jk}=v^jv^k,\ 1\leq j, k\leq 2,\ \ g_1^{j0}=g_1^{0j}=v^j,
\ 1\leq j\leq 2,\ g_1^{00}=0,
\\
\nonumber
&v=(v^1,v^2)=\frac{b_1(\theta)}{r}\hat r +\frac{b_2(\theta)}{r}\hat \theta,\ \ b_j,\ j=1,2, \ \ 
 b_1\neq 0,
\end{align} 
$g_2^{jk}$ are smooth in polar coordinates $(r,\theta),\ g_2^{00}\geq C>0,
\ g_2^{j0}=g_2^{0j}=O(r),\ \ 1\leq j\leq 2,\ [g_2^{jk}]_{j,k=1}^2$  is
 a negative definite matrix when $|x|<\e$,
\begin{equation}                                   \label{eq:3.4}
\big([g_2^{jk}]_{j,k=1}^2\alpha,\alpha\big)\leq -C_0|\alpha|^2,\ \ 
\forall\alpha=(\alpha_1,\alpha_2)\in \R^2.
\end{equation}
Writing the matrix $[g^{jk}]_{j,k=0}^2$ in polar coordinates we get
\begin{multline}                                \label{eq:3.5}
H\Big(r,\theta,\xi_0,\xi_r,\frac{\xi_\theta}{r}\Big)
=g^{00}\xi_0^2+2g^{r0}\xi_0\xi_r +2g^{\theta 0}\xi_0\frac{\xi_\theta}{r}
+g^{rr}\xi_r^2
+2g^{r\theta}\xi_r\frac{\xi_\theta}{r}
+g^{\theta\theta}\frac{\xi_\theta^2}{r^2}.
\end{multline}
Consider the system 
of null-bicharacteristics with the Hamiltonian (\ref{eq:3.5}):
\begin{equation}                             \label{eq:3.6}
\frac{dr}{ds}=
H_{\xi_r},
\frac{d\theta}{ds}=
H_{\xi_\theta},
\frac{dx_0}{ds}=
H_{\xi_0},
\frac{d\xi_r}{ds}=
-H_r,
\frac{d\xi_\theta}{ds}=
-H_\theta,
\frac{d\xi_0}{ds}=-H_{x_0}.
\end{equation}
We impose the following initial conditions:
\begin{equation}                                 \label{eq:3.7}
r(0)=\e,\ \ \theta(0)=\theta_0,\ \ \xi_r(0)=\eta_r,\ \ \xi_\theta(0)=\eta_\theta,      
\ \ \xi_0(0)=\eta_0.
\end{equation}
Since
$\frac{\partial H}{\partial x_0}=0$, we have that
$\xi_0(s)=\eta_0, \forall s$,   and we choose $\xi_0=\eta_0=0$.

There are two family of null-bicharacteristics when $\xi_0\equiv 0$:
\begin{equation}                                    \label{eq:3.8}
\xi_r^\pm(s)=\frac{-g^{r\theta}\pm\sqrt{(g^{r\theta})^2-g^{rr}g^{\theta\theta}}}{g^{rr}}\cdot
\frac{\xi_\theta^\pm(s)}{r}
=\frac{-g^{r\theta}\pm\sqrt{-\Delta}}{g^{rr}}\cdot\frac{\xi_\theta^\pm(s)}{r},
\end{equation}
assuming
that
$g^{rr}\neq 0$.

The decomposition (\ref{eq:3.2}) has the following form in polar coordinates
\begin{multline}                               \label{eq:3.9}
g^{rr}=\frac{b_1^2}{r^2}+g_2^{rr},\ \ g^{r\theta}=\frac{b_1b_2}{r^2}+g_2^{r\theta},\ \
g^{\theta\theta}=\frac{b_2^2}{r^2}+g_2^{\theta\theta},
\\
g^{r0}=\frac{b_1}{r}+g_2^{r0},\ \ g^{\theta 0}=\frac{b_2}{r}+g_2^{\theta 0}.
\end{multline}
\begin{theorem}                                \label{theo:3.1}
Suppose $b_1\neq 0$.  When $\e>0$  is small 
and $b_1<0$
 all null-bicharacteristics (\ref{eq:3.6}) starting at
$r=\e$ reach $r=0$  as $x_0$  increases.   
If $b_1>0$  then all null-bicharacteristics 
starting at $r=\e$ reach $r=0$ as $x_0$ decreases
\end{theorem}

{\bf Proof:}
Dividing  $\frac{dr}{ds}$  by $\frac{dx_0}{ds}$  and having $\xi_0=0$  we get from 
(\ref{eq:3.6}):
\begin{equation}                          \label{eq:3.10}
\frac{dr}{dx_0}=
\frac
{
\big(\frac{b_1^2}{r^2}+g_2^{rr}\big)\xi_r
+\big(\frac{b_1b_2}{r^2}+g_2^{r\theta}\big)\frac{\xi_\theta}{r}
}
{\big(\frac{b_1}{r}+g_2^{r0}\big)\xi_r
+\big(\frac{b_2}{r}+g_2^{\theta 0}\big)\frac{\xi_\theta}{r}}
\end{equation}
Substituting
(\ref{eq:3.8}) and (\ref{eq:3.9}) in (\ref{eq:3.10})  and cancelling 
$\frac{\xi_\theta^\pm}{r}$,  we get
\begin{multline}                   
\nonumber
\frac{dr^\pm}{dx_0}
=\frac{\big(\frac{b_1^2}{r^2}+O(1)\big)\big(-\frac{b_2}{b_1}\pm C_1r+O(r^2)\big)+
\frac{b_1b_2}{r^2}+O(1)}
{\big(\frac{b_1}{r}+O(r)\big)\big(-\frac{b_2}{b_1}\pm C_1r+O(r^2)\big)+
\frac{b_2}{r}+O(r)}
\\
=
\frac{\frac{\pm b_1^2}{r}C_1+O(1)}{\pm b_1C_1+O(r)}
=\frac{b_1}{r}+O(1).
\end{multline}
Since $b_1(\theta)\leq b_0<0$  we have $\frac{dr^\pm}{dx_0}<-\frac{|b_0|}{2r},$
for small $r$. Therefore
\begin{equation}
\nonumber
2r\frac{dr}{dx_0}<-|b_0|,\ \ (r^\pm(x_0))^2-(r^\pm(t))^2\leq
-|b_0|(x_0-t),\ \ r^\pm(t)>0.
\end{equation}
When $x_0$ increases 
$r^\pm(x_0)=0$  for some $x_0>t$.
When $b_1\geq b_0>0$  we have 
\begin{equation}
\nonumber
\frac{dr^\pm}{dx_0}\geq \frac{b_0}{2r},
\ \ 
(r^\pm(t))^2-(r^\pm(x_0))^2\geq b_0(t-x_0),
\end{equation}
When $x_0$  decreases we get $r^\pm(x_0)=0$  for some  $x_0<t$.
\qed

\subsection{Existence of black or white holes}
The following refinement of Theorem 4.1 in [E1] holds:
\begin{theorem}                            \label{theo:3.2}
Let $S$  be the ergosphere,  i.e. $S=\{x:\Delta(x)=0\}$ where 
$\Delta(x)=g^{11}g^{22}-(g^{12}(x))^2$.  We assume that $S$ is a smooth Jordan curve.
Let $\Omega$ be the interior of $S$ and let $O\in\Omega$.  Assume that $\Delta(x)<0$
for $x\in \Omega\setminus O$  and assume
 that $S$  is not characteristic at any $ x\in S$.
Assume that the metric tensor $[g^{jk}(x)]_{j,k=0}^2$  has a singularity at $O$
and conditions (\ref{eq:3.2}), (\ref{eq:3.3}),  (\ref{eq:3.4})
are satisfied.   If $b_1<0$  then there exists smooth Jordan curves $S^+$  and
$S^-$  in $\Omega$  both containing $O$,  $S^-$  is inside  the domain
with boundary $S^+$,  such that $S^+\times \R$  and $S^-\times\R$ are 
 black hole  horizons
and there is no  black hole  horizons between $S^+$  and $S$  and
between $S^-$  and $O$.

Analogously,  when $b_1>0$  there exists   white hole 
horizons $S^+\times\R$  and $S^-\times\R$.  When $S^-=S^+$ there is only one 
black hole or white hole horizon
between $S$ and $O$.
\end{theorem}

{\bf Proof:}
Let $S_\e=\{x: |x|=\e\}$,  $\e$ is small,  and $\Omega_\e=\Omega\setminus\{x: |x|<\e\}$.
Suppose $b_1<0$  (cf. (\ref{eq:3.3})).  It follows from Theorem \ref{theo:3.1}
that  both families of characteristics $x=x^+(x_0)$  and $x=x^-(x_0)$
(cf. (4.9) in [E1])  end on $S_\e$ when $x_0$  is increasing.
It was shown in [E1]  that
one of
these families,  say $x=x^+(x_0)$,  ends at $S$ and  the second  family  
$x=x^-(x_0)$  starts at $S$  when  $x_0$ increases.
 Therefore if some  curve $x=x_+^{(1)}(x_0)$  touches
$S_\e$  at some time $x_0^{(0)}$  it  can not
reach $S$  when $x_0<x_0^{(0)}$ decreases.  Therefore the limit set of 
the trajectory
$x=x_+^{(1)}(x_0)$ is contained inside $\Omega_\e$  and by the Poincare-Bendixson
theorem there exists a closed Jordan curve
$x=s_0(x_0)$  such that $S_0\times\R$  is the event horizon 
(i.e.  either black hole or white hole horizon),
where 
$S_0=\{x=s_0(x_0)\}$ (cf. Theorem 4.1 in [E1]).  
Consider any curve of the second family $x=x^-(x_0)$
on $S_0$.  If it is pointed 
inside $S_0$ then $S_0\times\R$  is a  black hole  horizon,  and there is
no  black hole  horizon inside $S_0$. 
We denote such $S_0$  by $S^-$.  If any curve $x=x^-(x_0)$  is pointed 
outward   of $S_0$ then $S_0\times\R$  is the white hole event horizon.
In this case consider any trajectory $x=x_-^{(1)}(x_0)$  of
$x=x^-(x_0)$  family  that reaches $S_\e$  at some time $x_0^{(1)}$.
When $x_0<x_0^{(1)}$  decreases $x=x_-^{(1)}(x_0)$  can not reach $S_0$ and
again  by the Poincare-Bendixson theorem there exists closed Jordan
characteristic curve $x=s_1(x_0)$  such that $S^{(1)}\times\R$  is a  black 
hole  horizon  where $S^{(1)}=\{x=s_1(x_0)\}$ and there are no 
black holes
inside $S^{(1)}$.  In this case we set $S^{(1)}=S^-$.   Therefore in
both cases $S^-\times\R$  is the smallest 
 black hole  horizon in $\Omega\setminus O$.

Consider now a curve $x=x_+^{(2)}(x_0)$ of the $x=x^+(x_0)$  family that  ends
at $S$  at the time $x_0=x_0^{(2)}$.  When $x_0<x_0^{(2)}$  
decreases this trajectory can not reach $S_\e$ and by 
the Poincare-Bendixson theorem there is a closed Jordan 
trajectory $x=s_2(x_0)$ such that $S_2=\{x=s_2(x_0)\}\times\R$  is 
an event horizon. If $S_2\times\R$ is a  black hole  horizon we denote
$S_2$  by $S^+$.  If $S_2\times\R$  is a  white hole  horizon then we have, as above,
that there is a closed characteristic curve  $S_3=\{x=s_3(x_0)\}$  belonging to 
$x=x^-(x_0)$  
family such that $S_3\times\R$  is a  black hole  horizon and
there are no  black hole  horizons between 
$S_3$ and $S$.
In this  case we denote $S_3$   by $S^+$.

This concludes the proof of  Theorem \ref{theo:3.2}
in the case $b_1<0$.  If 
$b_1>0$ then the same reasoning leads to the proof that there are 
two  white hole  horizons $S^+\times\R$
and $S^-\times\R$  such that
there is no white hole  horizons between $S^+$
and $S$  and between $S^-$  and $O$.  In the case when $S^-=S^+$  we have
a unique outer black or white hole in $\Omega\setminus O$.
\qed

Note the 
following property of the black hole with the boundary 
$S^+\times\R$:  disturbance at any point 
outside $S^+\times\R$
propagate to the spatial
infinity.  The black hole  horizon 
$S^-\times\R$  has the property that any  point inside $S^-\times\R$  ends on $O$.

In Example 4.2  in [E1]  the case when there are many 
black 
hole horizons and 
white hole horizons was considered.

{\bf Remark 3.1}
In the  paper [EH] by Michael Hall and the author 
(see also Michael Hall, PhD dissertation, UCLA, 2013)  
the condition that the ergosphere is
not characteristic at any point is removed.
\section{Spherically symmetric black holes for the nonstationary metrics}
\init
\subsection{Spherically symmetric black holes}
Let the nonstationary  metric tensor 
$[g_{jk}(x_0,x)]_{j,k=0}^n$  be the same
as in (\ref{eq:1.1}).  We assume that  the metric has a singularity  at
$O\times\R$  that satisfies (\ref{eq:3.2}),  (\ref{eq:3.3}),
(\ref{eq:3.4})  uniformly in $x_0$.  We shall  introduce spherical coordinates
$(r,\theta)$ in $\R^n$ and we shall write the tensor $[g^{jk}]_{j,k=0}^n$
in such coordinates (cf.  (\ref{eq:3.5})  for the  
case of $n=2$).
We shall not assume that the metric is spherically
symmetric but we shall require 
that $g^{00}(x_0,r),\ g^{0r}=g^{r0}(x_0,r)$  and $g^{rr}(x_0,r)$  be 
dependent of $x_0,r=|x|$  only.

We are looking for the characteristic
surface of the form $S(x_0,r)=0$.  Then the characteristic equation has
the form
\begin{equation}                                        \label{eq:4.1}
g^{00}(x_0,r)S_{x_0}^2+2g^{r0}(x_0,r)S_{x_0}S_r+g^{rr}(x_0,r)S_r^2=0
\end{equation}
when  $S(x_0,r)=0$.

Note that
$S(x_0,r)=0$  is a curve in $(x_0,r)$  space but it is a $(n-1)+1$ dimensional
surface in $\R^n\times\R$.  We shall study when $S(x_0,r)=0$ is a 
black hole
horizon starting with the case of nonstationary acoustic metric.
\subsection{Nonstationary acoustic metric}
Consider a time-dependent acoustic metric (\ref{eq:3.1})
in $(\R^2\setminus O)\times\R$  with the velocity flow of the form
\begin{equation}                                  \label{eq:4.2}
v=\frac{A(x_0)}{r}\hat r+\frac{B(x_0,r,\theta)}{r}\hat\theta.
\end{equation}
The characteristic equation 
for $S(x_0,r)=0$ is
$$
\Big(S_{x_0}(x_0,r)+\frac{A(x_0)}{r}S_r\Big)^2-S_r^2=0,
$$
or
\begin{equation}                               \label{eq:4.3}
S_{x_0}^\pm=\Big(-\frac{A(x_0)}{r}\pm 1\Big)S_r^\pm \ \ \mbox{when}\ \ 
S(x_0,r)=0.
\end{equation}
Assuming
that $S_r\neq 0$  when  $S=0$ we can, without loss of generality,  look
for $S$ in the form $S=r-r(x_0)=0$.  Then $r=r(x_0)$  satisfies 
the ordinary differential equation
\begin{equation}                         \label{eq:4.4}
r_{x_0}(x_0)=\frac{A(x_0)}{r}\mp 1.
\end{equation}
There are two family of solutions $r=r^+(x_0)$  and $r=r^-(x_0)$  satisfying
the equations $r_{x_0}^+=\frac{A(x_0)}{r}+1$ and
$r_{x_0}^-=\frac{A(x_0)}{r}-1$, respectively.  We assume $A(x_0)\rw A(\pm\infty)$ 
when $x_0\rw \pm\infty$.  We shall show that there exists a unique (outer)
black hole horizon if $A(x_0)\leq A_0<0$  and  
there exists a unique (outer)
white hole horizon if $A(x_0)\geq A_0>0$.  When $A(x_0)=A_0$  there is an explicit 
solution of $\frac{dr^\pm}{dx_0}=\frac{A_0}{r}\mp 1$.  In particular,
when $A_0<0$  and $\frac{dr}{dx_0}=-\frac{|A_0|}{r}+1$.  
We have
\begin{equation}                                     \label{eq:4.5}
\frac{r}{r-|A_0|}dr=dx_0,\ \ r+|A_0|\ln|r-|A_0||=x_0+C,
\end{equation}
$r^+(x_0)\rw |A_0|$  when $x_0\rw -\infty$.
If $r^+(0)=r_0<|A_0|$
then $r^+(x_0)$  decreases to 0 when $x_0$ increases.
If $r^+(0)=r_0>|A_0|$
  then  $r^+(x_0)\rw +\infty$  when
$x_0\rw +\infty$ (cf.  Fig. 1a).
The (outer) black hole horizon is $\{r=|A_0|\}\times\R$.  The solution $r^+=|A_0|$  is 
the separatrix separating the solutions tending to $+\infty$ 
when $x_0\rw +\infty$  and the solutions ending on $r=0$  at some $x_0=x_0^{(0)}$.

\makefig{\begin{minipage}{5.0in}
\smallskip
\begin{itemize}
\item[{\rm (1a)}] The family $r=r^+(x_0)$,\ $A(x_0)=A_0<0$.
$\{r<|A_0|\}\times\R$  is a black hole.
\item[{\rm (1b)}] The family $r=r^-(x_0),\ A(x_0)=A_0<0$.
\end{itemize}
\end{minipage}}
{fig:1a1b}{\includegraphics{Figures1a1b.eps}}

\makefig{\begin{minipage}{5.0in}
\smallskip
\begin{itemize}
\item[{\rm (2a)}] Trajectories $r=r^+(x_0)$  when  $A_0>0.$
\item[{\rm (2b)}] Trajectories $r=r^-(x_0)$ when $A_0>0$.  $\{r<A_0\}\times\R$ is a white hole.
\end{itemize}
\end{minipage}}
{fig:2a2b}{\includegraphics{Figures2a2b_new.eps}}

Since $\frac{dr^-(x_0)}{dx_0}=-\frac{|A_0|}{r}-1<-1$  all trajectories 
$r=r^-(x_0)$
end on $r=0$  when $x_0$  increases (see Fig. 1b).  Analogously,  if 
$A(x_0)=A_0>0$  then  $\{r<A\}\times\R$  is a white hole.  All trajectories of 
$\frac{dr^+(x_0)}{dx_0}=\frac{A_0}{r}+1>1$ 
start at $r=0$ and tend  to $+\infty$  when $x_0$  increases (see Fig. 2a).
For the trajectories of $\frac{dr^-}{dx_0}=\frac{A_0}{r}-1$  we have that the solution
$r=A_0$  separates the solutions that tend to 0 and to $+\infty$  when $x_0$  decreases.

Now consider the case of $A(x_0)$  depending on $x_0$.
Let $A(x_0)\leq -A_0<0$.  Since $\frac{dr^-}{dx_0}=-\frac{|A(x_0)|}{r}-1<-1$ all
trajectories $r=r^-(x_0)$ end
on $r=0$  (cf. Fig 1b.).  Consider 
\begin{equation}                                \label{eq:4.6}
\frac{dr^+(x_0)}{dx_0}=\frac{A(x_0)}{r}+1=\frac{r+A(x_0)}{r}.
\end{equation}
Make change of variables
$r^++A(x_0)=v(x_0)$,
where $v(x_0)$ will be chosen later.  Then
$\frac{dr^+}{dx_0}=\frac{dv}{dx_0}-A'(x_0)$
and $v(x_0)$  satisfies the equation
\begin{equation}                              \label{eq:4.7}
\frac{dv}{dx_0}-\frac{v}{|A(x_0)|+v}=A'(x_0).
\end{equation}
Let $w=v\exp(-b(x_0))$,  where $b(x_0)=\int_0^{x_0}\frac{dy_0}{|A(y_0)|+v(y_0)}.$

Then 
$
\frac{dw}{dx_0}\exp b(x_0)=A'(x_0)
$
and
$
w(x_0)=w(0)+\int_0^{x_0}A'(t)\exp(-b(t))dt.
$

Therefore $v=\Big(w(0)+\int_0^{x_0}A'(t)\exp(-b(t))dt\Big)\exp b(x_0)$.
Choosing $w(0)=-\int_0^\infty A'(t) \exp (-b(t))dt$  we have the following
representation for particular solution $v(x_0)$ of (\ref{eq:4.7}) 
\begin{equation}                            \label{eq:4.8}
v(x_0)=-\int_{x_0}^\infty 
A'(t)\exp\Big(-\int_{x_0}^t\frac{dy_0}{|A(y_0)|+v(y_0)}\Big)dt.
\end{equation}
Note  that  $v(x_0)\rw 0$ when $x_0\rw +\infty$  since  $t>x_0$  in (\ref{eq:4.8}).
Denote 
by $B$  the Banach space with norm 
 $\|v\|_T=\sup_{x_0\geq T}|v(x_0)|,\ \ T$ will be large enough. 
Denote by $F(v)$  the operator
\begin{equation}                            \label{eq:4.9}
F(v)=-\int_{x_0}^\infty 
A'(t)\exp\Big(-\int_{x_0}^t\frac{dy_0}{|A(y_0)|+v(y_0)}\Big)dt.
\end{equation}
We have when, $\|v\|_T<\frac{1}{2}|A_0|$,
\begin{equation}                             \label{eq:4.10}
\|F(v)\|_T\leq\int_T^\infty|A'(t)|dt.
\end{equation}
Also  we get, when $\|v_j\|_T<\frac{|A_0|}{2},\j=1,2,$
\begin{equation}                           \label{eq:4.11}
\|F(v_1)-F(v_0)\|_T\leq  C\int_T^\infty t|A'(t)|dt\|v_1-v_2\|_T.
\end{equation}
Therefore  $F(v)$ is a contraction map  of a ball $\|v\|_T\leq \e$
if $T$  is sufficiently large.  Hence there exists a unique bounded solution
$r_0^+(x_0)=-A(x_0)+v$  of (\ref{eq:4.6}) for $x_0>T$.  Taking in (\ref{eq:4.8})
the limit when $x_0\rw +\infty$  we get
$\lim_{x_0\rw +\infty}r_0^+(x_0)=-\lim_{x_0\rw +\infty}A(x_0)+
\lim_{x_0\rw +\infty}v(x_0)=-A(+\infty)$.
Since $A(x_0)\leq -A_0<0$  for all $-\infty<x_0<+\infty$  and 
$\frac{dr_0^+(x_0)}{dx_0}=\frac{A(x_0)}{r}+1$  we can extend  
$r=r_0^+(x_0)$ from $(T,+\infty)$  to $(-\infty,+\infty)$:

For small $r$  we have $\frac{dr_0^+}{dx_0}<-\frac{|A_0|}{2r_0}$.  Thus when 
$r_0^+(x_0)$ is small it increases when $x_0$  decreases and hence it cannot
hit $r=0$  for any  
$x_0\in\R$.  If $r$  is large then $\frac{dr_0^+}{dx_0}>\frac{1}{2}$  and 
$r_0^+(x_0)$  decreases  when  $x_0$  decreases.  Therefore 
$r=r_0^+(x_0)$  will remain bounded 
and nonzero for all $x_0\in \R$.
Since  $b(x_0)<0$  when $x_0<0$  and $|v(x_0)|<\frac{1}{2}|A_0|$  all solutions of
(\ref{eq:4.7})  tend  to zero  when  $x_0\rw -\infty$.  Therefore
$\lim_{x_0\rw -\infty}r_0^+(x_0)=\lim_{x_0\rw -\infty}-A(x_0)+
\lim_{x_0\rw -\infty}v(x_0)=-A(-\infty)$.

Therefore we proved the following theorem.
\begin{theorem}                                       \label{theo:4.1}
Consider the acoustic  metric with the velocity  
flow (\ref{eq:4.2}).  Assume that $A(x_0)\in C^\infty(\R),\ A(x_0)\leq A_0<0,\ 
\lim_{x_0\rw\pm\infty}A(x_0)
=A(\pm\infty)$. Then there exists  a unique (outer)
black hole horizon $r=r_0^+(x_0)$  and $\lim_{x_0\rw \pm\infty}r_0^+(x_0)=
|A(\pm\infty)|$.
The solution $r=r_0^+(x_0)$  of (\ref{eq:4.6}) separates the solutions
 $r=r^+(x_0)$  that reach $r=0$  at some time $x_0$ and the solutions
$r=r^+(x_0)$  that tend to the infinity when $x_0\rw +\infty$.
Analogously when $A(x_0)\geq A_0>0$   there exists a unique (outer)
white hole horizon $r=r_0^-(x_0)$  such that $\lim_{x_0\rw\pm\infty}r_0^-(x_0)=
A(\pm\infty)$.
\end{theorem}
The solution $r=r^-(x_0)$  separates  the  solutions  $r=r^-(x_0)$ that tend  to $\infty$
when $x_0\rw -\infty$  and the solutions  $r=r^-(x_0)$  that  end  on  $r=0$  when  $x_0$ 
decreases.

\subsection{Appearance of black holes at finite time}
Suppose $A(x_0)$  in (\ref{eq:4.6})
is such that $A(x_0)<0$  when $x_0>0$ and $A(x_0)\geq 0$  for $x_0<0$.
It was proven above  that there exists a solution $r=r_0^+(x_0)$  of (\ref{eq:4.6})
for $x_0>T,\ T$ is large.  We shall show  that this solution will vanish  at some
point $x_0^{(1)}<0$.  Let  $x_0<0$.   Then 
$\frac{dr_0^+(x_0)}{dx_0}=\frac{A(r_0)}{r}+1>1$  since  $A(x_0)\geq 0$  for $x_0<0$.
Thus $r_0^+(x_0)$  decreases when $x_0$  decreases and we have, at some time 
$x_0=x_0^{(1)}$, that $r_0^+(x_0^{(1)})=0$.   Therefore  the black 
hole $\{r<r_0^+(x_0)\}\times(x_0^{(1)}<x_0<+\infty)$  starts  at the time
$x_0^{(1)}$  and there is no black hole for $x_0<x_0^{(1)}$.

Analogously suppose  in (\ref{eq:4.6})  we have $A(x_0)>0$  for
$x_0<0$  and $A(x_0)< 0$  for  $x_0>0$.  Let  $r=r_0^-(x_0)$  be 
the white hole   horizon on $(-\infty,-T)$.   When  $x_0>0$  we have 
$\frac{dr_0^-}{dx_0}=\frac{A(x_0)}{r}-1\leq -1$  since 
$A(x_0)<0$  for $x_0>0$. Therefore $r^-(x_0^{(2)})=0$
for some $x_0^{(2)}$.  This means  that 
the white hole
$\{r<r_0^-(x_0)\}\times(-\infty,x_0^{(2)})$  ends at the time $x_0=x_0^{(2)}$.

\subsection{The general case}
Consider the metric such that $g^{00},g^{r0},g^{rr}$  depend on 
$(x_0,r)$  only so that the characteristic equation has the form (\ref{eq:4.1}).
We assume that $g^{00}(x_0,r)\geq C_0>0$.  We have
\begin{equation}                                   \label{eq:4.12}
S_{x_0}^\pm=\frac{
-g^{r0}(x_0,r)\pm \sqrt{
(g^{r0})^2-g^{00}g^{rr}
}
}
{g^{00}}
S_r^\pm.
\end{equation}
Note that $S_{x_0}^+$  is the larger root in (\ref{eq:4.12}) and
$S_{x_0}^-$ is the smaller root.  

The strict hyperbolicity implies that
\begin{equation}                           \label{eq:4.13}
(g^{r0})^2-g^{00}g^{rr}\geq C_0>0\ \ \forall x_0,r.
\end{equation}
We shall assume that 
\begin{equation}                               \label{eq:4.14}
g^{00}=1+O\Big(\frac{1}{r}\Big),\ \ g^{r0}=\Big(\frac{1}{r}\Big),\ \
g^{rr}=-1+O\Big(\frac{1}{r}\Big)
\end{equation}
when $r$  is large.
Near $r=0$ we assume (cf. (\ref{eq:3.2}), (\ref{eq:3.3}), (\ref{eq:3.4}))
\begin{equation}                                 \label{eq:4.15}
g^{r0}=\frac{b_1(x_0)}{r}+g_2^{r0},\ \ g^{rr}=\frac{b_1^2(x_0)}{r^2}+g_2^{rr},\ \ 
g^{00}=1+g_2^{00},
\end{equation}
where 
$g_2^{r0}=O(r^2),\ g_2^{00}=O(r^3),\ g_2^{rr}\leq -C_0<0, \ b_1(x_0)\leq b_0<0$.
Then
$$
(g^{r0})^2-g^{00}g^{rr}=-g_2^{rr}+O(r)\geq \frac{C_0}{2}
$$ 
for  $r$ small uniformly in $x_0$.
Therefore near $r=0$  we have
\begin{equation}                             \label{eq:4.16}
S_{x_0}^\pm=\Big(-\frac{b_1(x_0)}{r}\pm g_0(x_0,r)\Big)S_r^\pm,
\end{equation}
where  $g_0\geq C>0.$

Looking for $S^\pm=0$ in the form $S^\pm=r-r^\mp(x_0)=0$  we get ordinary
differential equations for $r=r^\pm(x_0)$
\begin{equation}                              \label{eq:4.17}
 r_{x_0}^+=\frac{g^{r0}(x_0,r)}{g^{00}(x_0,r)}+
d(x_0,r),
\ \ \ \
 r_{x_0}^-=\frac{g^{r0}(x_0,r)}{g^{00}(x_0,r)}-
d(x_0,r),
\end{equation}
where
$d(x_0,r)\geq C_0$.  Since $S^\pm=r-r^\mp(x_0)$  we have that $r^+$ corresponds 
to the smaller root in (\ref{eq:4.12})  and $r^-(x_0)$  corresponds  to the 
larger root in (\ref{eq:4.12}).

When $r$  is close to 0  we have
\begin{equation}                               \label{eq:4.18}
r_{x_0}^\pm=\frac{b_1(x_0)}{r}\pm d_1(x_0,r),\ \ d_1\geq C_1>0.
\end{equation}
When $r$  is large  we have
\begin{equation}                               \label{eq:4.19}
r_{x_0}^\pm=\pm 1 + O\Big(\frac{1}{r}\Big).
\end{equation}
Suppose 
$b_1(x_0)<b_0<0$  for all $x_0\in\R$.  It follows from (\ref{eq:4.18}) that
\begin{equation}                              \label{eq:4.20}
\frac{dr^\pm(x_0)}{dx_0}<-\frac{1}{2}|b_0| \ \ \mbox{for small}\ \ r.
\end{equation}
Therefore
if $r^\pm(x_0^{(1)})=r_0$  is small then 
$(r^\pm(x_0))^2-(r^\pm(x_0^{(1)}))^2\leq -\frac{1}{2} |b_0|(x_0-x_0^{(1)})$  and
$r^\pm(x_0^{(2)})=0$ for some $x_0^{(2)}>x_0^{(1)}$.
If $r^+(x_0^{(1)})=r_0$  is large  it follows from (\ref{eq:4.19}) that 
$r^+(x_0)\rw +\infty$  since
\begin{equation}                                \label{eq:4.21}
\frac{dr^+(x_0)}{dx_0}>\frac{1}{2}\ \ \mbox{for}\ \ r^+(x_0)>r_0.
\end{equation} 
Denote by $\mathcal F^+$ the set of all $r^+(x_0)$  that tend 
to $+\infty$  when $x_0\rw +\infty$.  For each $t\in\R$ define  by $R^+(t)$  the
infimum of $r^+(t)$ 
over all $r^+(x_0)\in\mathcal F^+$: 
$
R^+(t)=\inf_{\mathcal F^+} r^+(t).
$

Note that
$R^+(t)>0$:
if $r^+(t)$  is small then $r^+(t)\not\in\mathcal F^+$  
since
it will end at $r=0$.  Let  $R^+(x_0)$  be the solution 
of of (\ref{eq:4.17})  passing through $R^+(t)$  at $x_0=t$.  Since
$R^+(t)\leq r^+(t),\ \forall r^+\in\mathcal F^+$,  we have, for
any $x_0$,  that $R^+(x_0)\leq r^+(x_0),\ \forall r^+\in \mathcal F^+$.
The solution $R^+(x_0)$  is bounded when $x_0\rw +\infty$  since  if $R^+(x_0)$
is unbounded  then  we will have $\frac{dR^+}{dx_0}>\frac{1}{2}$
for $x_0\geq x_0^{(0)}$.  Then, choosing
$r_0$  such that
 $R^+(x_0^{(0)})-\e<r_0<R^+(x_0^{(0)})$,
where  $\e$  small,  we get that the solution
$r_0^+(x_0)$,  where $r_0^+(x_0^{(0)})=r_0$,  also tends to $+\infty$  when
$x_0\rw +\infty$,  i.e. $r_0^+\in \mathcal F^+$.  Since  $r_0^+(x_0^{(0)})<
R^+(x_0^{(0)})$,  we have  that $r_0^+(x_0)<R^+(x_0)$  for all $x_0$.
In particular,  $r_0^+(t)<R^+(t)$  and
this is a contradiction,  since
   $R^+(t)=\inf_{\mathcal F^+}r^+(t)$. 
Note that the solution  $R^+(x_0)$  exists for all $x_0\in\R$:
it can not  end at $r=0$ since $b_1(x_0)<0,\ \forall x_0$,
and for small $r$,\ $r^+(x_0)$  increases when $x_0$ decreases  (cf.  (\ref{eq:4.20})),
and it can not escape to the infinity when 
$x_0\rw -\infty$  since, for large $r$,  $r^+(x_0)$  is decreasing  when  $x_0$  is 
decreasing  (see  (\ref{eq:4.21})).  Note
that $R^+(t_1)=\inf_{r^+\in \mathcal F}r^+(t_1)$  for any $t_1$:
Suppose  $R_1^+(t_1)=\inf_{\mathcal F^+}r^+(t_1)>R^+(t_1)$.  Denote
by $R_1^+(x_0)$  the solution of (\ref{eq:4.17})    passing through
$R_1^+(t_1)$ when  $x_0=t_1$.  Since $R_1^+(t_1)>R^+(t_1)$  we have that
$R_1^+(t)>R^+(t)$  and this contradicts that $R^+(t)=\inf_{\inf_{\mathcal F^+}} r^+(t)$,
since
$r^+(t)\geq R_1^+(t)>R^+(t)$  for $\forall r^+\in \mathcal F^+$.

Since $R^+(x_0)$ corresponds to the smaller root $S^-$  in (\ref{eq:4.12})
the solution $r=R^+(x_0)$  is an outer black hole horizon (cf. Definition 1.2).
Now denote by $\mathcal F^-$  the set of  solutions $r=r^+(x_0)$ that
end at $r=0$  at some time $x_0=x_0^{(0)}$.
We extend $r=r^+(x_0)$  by zero  for $x_0\geq x_0^{(0)}$.
For fixed $t\in\R$  denote  by $R_+(t)$ the supremum of $r^+(t)$,
where $r^+\in \mathcal F^-: R_+(t)=\sup_{\mathcal F^-}r^+(t)$.
Let $R_+(x_0)$  be a solution of (\ref{eq:4.17})  
 passing through $R_+(t)$  at $x_0=t$.  Note that $R_+(t)<+\infty$  
since any $r^+(x_0)\in \mathcal F^+$  is
larger than any $r^+(x_0)\in \mathcal F^-$.
The same arguments as for  $R^+(x_0)$  
show  that  $R_+(t_1)=\sup_{\mathcal F^-}r^+(t)$  
for any $t_1$.

As for $R^+(x_0)$  we have that 
$R_+(x_0)$  exists for all $x_0\in \R$  providing that $b(x_0)\leq b_0<0$
for all $x_0\in \R$.
Therefore
$R_+(x_0)$  is a  black hole horizon.  
Note that $R_+(x_0)\leq R^+(x_0),\ \forall x_0$.

We proved the following theorem:
\begin{theorem}                            \label{theo:4.2}
Consider a nonstationary metric such that $g^{00}, g^{r0}, g^{rr}$
depend on $(x_0,r)$  only.  Let  
$g^{00}(x_0,r)\geq C_0>0$.
Assume  that conditions (\ref{eq:4.13}),  (\ref{eq:4.14}), (\ref{eq:4.15})
are satisfied.  Suppose $b_1(x_0)\leq b_0<0$  in (\ref{eq:4.15}),  $\forall x_0\in \R$.
Then  there exists an outer  black hole
horizon  $r=R^+(x_0)$ such that  $R^+(x_0)>0,\ \forall x_0\in\R,\ \ R^+(x_0)$   
is bounded on $\R$.\linebreak   Moreover,  $r=R^+(x_0)$  is the boundary of the set
of all solutions $r=r^+(x_0)$  of (\ref{eq:4.17})
that tend to the infinity when $x_0\rw +\infty$.  Also there exists a
 black  hole horizon $r=R_+(x_0)$
such that $R_+(x_0)>0$  on $\R,\ R_+(x_0)$  is bounded on $\R,\ R_+(x_0)\leq R^+(x_0)$  and  $R_+(x_0)$ is the boundary  of all solutions of (\ref{eq:4.17})  that end at
$r=0$  for  some $x_0\in \R$.  If $R_+(x_0)=R^+(x_0)$  there is a 
unique black hole horizon.
\end{theorem}

{\bf Remark 4.1}
Analogous results hold  for  white hole horizons assuming that 
$b_1(x_0)\geq b_0>0$
for all $x_0$.  In particular,  there exists   white hole 
horizons $r=R^-(x_0)$  and  $r=R_-(x_0)$  such that $R^-(x_0)$  is
the boundary of the set of  all solutions $r=r^-(x_0)$
such that  $r^-(x_0)\rw +\infty$ when 
$x_0\rw -\infty$  and $r=R_-(x_0)$  is the boundary of the set 
 of all $r=r^-(x_0)$
such that $r^-(x_0)$ ends at $r=0$  when $x_0$  decreases.

\section{Inverse problems and black or white holes}
\init

Consider nonstationary  spherically symmetric metric
\begin{equation}                                 \label{eq:5.1}
g_{00}(x_0,r)dx_0^2+2g_{0r}(x_0,r)dx_0dr+g_{rr}'(x_0.r)dr^2-dr^2-r^2(d\theta)^2,
\end{equation}
where $(d\theta)^2$  is a standard metric on $S^{n-1}$.

Let
$$
\Box_g u(x,t)=0
$$
be the wave equation  of the form  (\ref{eq:1.2})  corresponding to the metric 
(\ref{eq:5.1}).  Suppose $g_{00}, g_{0r}$  and $g_{rr}=-1+g_{rr}'$  satisfy the condition
of the Theorem \ref{theo:4.2}.

Let $r=R^+(x_0)$  be the boundary of the outer black hole, i.e. $b_1(x_0)<0$  
in the condition (\ref{eq:4.15}).
Let $C$ be
 a cylinder $C=\{r<a,-\infty<x_0<+\infty\}$  such  that $r=R^+(x_0)$  is inside
$C$,  i.e.  $a$  is
large enough.

Consider the following  initial-boundary value problem
\begin{equation}                                  \label{eq:5.2}
\Box_gu=0\ \ \mbox{in}\  \ C,
\end{equation}
\begin{equation}                                 \label{eq:5.3}
u=0\ \ \mbox{for}\ \ x_0\ll 0,
\end{equation}
\begin{equation}                                  \label{eq:5.4}
u\Big|_{\partial C}=f,\ \ f\in C_0^\infty(\partial C).
\end{equation}
Let $\Lambda f$  be the Dirichlet-to-Neumann operator,  i.e.  
$\Lambda f=\frac{\partial u}{\partial\nu}\big|_{\partial C}$
where $\nu$  is the outward  unit  normal to $\partial C$,  $u$  is the solution  of 
(\ref{eq:5.2}), (\ref{eq:5.3}), (\ref{eq:5.4}).
The inverse hyperbolic boundary  value problem  consists of recovery of the metric
knowing  $\Lambda f$  on  $\partial C$  for all $f\in C_0^\infty(\partial C)$,  i.e.
knowing  the Cauchy data  of $u$  on $\partial C$.
It is impossible to recover  the metric inside the black  hole since no information  from the
interior of the black hole  can reach  $\partial C$.  Therefore changes of the metric 
inside $r<R^+(x_0)$  does not effect  boundary data on $\partial C$.
However,  we shall prove that knowing  the Cauchy data on $\partial C$  one can 
 recover  the metric outside the black hole,  i.e.  when $r> R^+(x_0)$.
Assuming that $f$ does not  depend on $\theta$  on $\partial C$  we shall consider  the
solution of the wave equation  
depending on $(x_0,r)$  only 
\begin{align}                                            \label{eq:5.5}
&\frac{1}{\sqrt{- g(x_0,r)}}\frac{\partial}{\partial x_0}
\Big(\sqrt{-g}\, g^{00}(x_0,r)\frac{\partial u}{\partial x_0}\Big)
+\frac{1}{\sqrt{-g}}\frac{\partial}{\partial x_0}
\Big(\sqrt{-g}\,g^{0r}\frac{\partial u}{\partial r}\Big)
\\
\nonumber
+&\frac{1}{\sqrt{-g}}\frac{\partial}{\partial r}\Big(\sqrt g\, g^{0r}
\frac{\partial u}{\partial x_0}
\Big)
+\frac{1}{\sqrt{-g}}\frac{\partial}{\partial r}\Big(\sqrt{-g}\,g^{rr}
\frac{\partial u}{\partial r}\Big)
=0,\ r>0, \ x_0\in \R,
\end{align}
where $g^{00},g^{0r}=g^{r0}, g^{rr}$  are  the same as in (\ref{eq:4.1}).
Let  $\varphi(x_0,r)$  be 
a solution of
\begin{equation}                                 \label{eq:5.6}
g^{00}\varphi_{x_0}^2+2g^{r0}\varphi_{x_0}\varphi_r+g^{rr}\varphi_r^2=0,\ \ r>0.
\end{equation}
We assume,  as in \S4,  that $g^{00}\geq C>0$  and  conditions 
(\ref{eq:4.13}),  (\ref{eq:4.14}), (\ref{eq:4.15}) are satisfied.

Note that we can factor (\ref{eq:5.6}):
\begin{equation}                                     \label{eq:5.8}
g^{00}\varphi_{x_0}+g^{0r}\varphi_r\mp\sqrt{(g^{0r})^2-g^{00}g^{rr}}\varphi_r=0.
\end{equation}
Denote  by $\varphi_1(x_0,r),\varphi_2(x_0,r)$  two solutions  of (\ref{eq:5.8})
\begin{align}                                 \label{eq:5.9}
g^{00}\varphi_{1x_0}+b_-(x_0,r)\varphi_{1r}=0,
\\
g^{00}\varphi_{2x_0}+b_+(x_0,r)\varphi_{2r}=0, \label{eq:5.10}
\end{align}
where
\begin{equation}                                \label{eq:5.11}
b_\pm(x_0,r)=g^{0r}\pm\sqrt{(g^{0r})^2-g^{00}g^{rr}}.
\end{equation}
Note  that  $b_\pm(x_0,r)\neq 0$  near  $r=a$.
Make  change of variables 
\begin{equation}                           \label{eq:5.12}
s=\varphi_1(x_0,r),\ \ \tau=\varphi_2(x_0,r).
\end{equation}
Then (\ref{eq:5.5}) has the following form in $(s,\tau)$  coordinates:
\begin{equation}                                     \label{eq:5.13}
\hat g^{s\tau}\frac{\partial^2\hat u}{\partial s\partial \tau}=0,
\end{equation}
where 
\begin{equation}                                   \label{eq:5.14}
\hat g^{s\tau}=g^{00}\varphi_{1x_0}\varphi_{2x_0}+
g^{r0}(\varphi_{1x_0}\varphi_{2r}+\varphi_{1r}\varphi_{2x_0})+
g^{rr}\varphi_{1r}\varphi_{2r}.
\end{equation}
We used that (\ref{eq:5.9}), (\ref{eq:5.10})  imply  that
$\hat g^{ss}=\hat g^{\tau\tau}=0$  and that 
$\sqrt{-\hat g}=(\hat g^{s\tau})^{-1}.$

We impose the initial conditions on $\varphi_1,\varphi_2$:
\begin{equation}                                  \label{eq:5.15}
\varphi_1\big|_{r=a}=x_0+a,\ \ \ \ \ \varphi_2\big|_{r=a}=-x_0+a.
\end{equation}
Let
\begin{equation}                                      \label{eq:5.16}
y_0=\frac{s-\tau}{2}=\frac{\varphi_1(x_0,r)-\varphi_2(x_0,r)}{2},
\ \ \
y_1=\frac{s+\tau}{2}=\frac{\varphi_1(x_0,r)+\varphi_2(x_0,r)}{2}
\end{equation}                       
Then 
\begin{equation}                                         \label{eq:5.17}
y_0\big|_{r=a}=\frac{x_0-(-x_0)}{2}=x_0,
\ \ \
y_1\big|_{r=a}=\frac{x_0+a+(-x_0+a)}{2}=a
\end{equation}
Therefore the map $(x_0,r)\rw(y_0,y_1)$  is the identity  when   $r=a$.

Making the changes of variables  (\ref{eq:5.16})  we get  from  (\ref{eq:5.13})
\begin{equation}                                         \label{eq:5.18}
\frac{\partial^2\hat u(y_0,y_1)}{\partial y_0^2}-\frac{\partial^2\hat u(y_0,y_1)}{\partial y_1^2}
=0,
\end{equation}
where
\begin{equation}                                       \label{eq:5.19}                
\hat u(y_0,y_1)=u(x_0,r),
\end{equation}
$(y_0,y_1)$ and  $(x_0,r)$  are related by the equations (\ref{eq:5.16}).

The characteristics of (\ref{eq:5.5}),  crossing the line $r=a$,  have the form
\begin{equation}                              \label{eq:5.19}
r-r^+(x_0,C_1)=0,\ \ \ r-r^-(x_0,C_2)=0,
\end{equation}
where
$
\frac{dr^\pm}{dx_0}=\frac{b_\mp(x_0,r)}{g^{00}(x_0,r)}.
$

As in \S 4 solutions $r=r^+(x_0,C_1),\ \forall C_1,$  
tend  to $+\infty$  when $x_0\rw+\infty$,  and $r=r^-(x_0,C_2)$  end at  $r=0$  when
$x_0$  increases,  $\forall C_2$.
It was shown  in Theorem  \ref{theo:4.2} that  $r=R^+(t)=\inf_{C_1}r^+(x_0,C_1)$ 
 is the black hole horizon.

The images of the characteristics (5.19) under the map (\ref{eq:5.16}) are
the characteristics
$
s=C_1,\  \tau=C_2
$
of  the equation (\ref{eq:5.18}).

Denote by  $D_g$  the subdomain  in $[0,a]\times\R^1$  bounded by the  equation
$r=R^+(x_0)$  and the  line $\{r=a,x_0\in \R\}$.  Then  (\ref{eq:5.16}) maps
$D_g$   onto  the half-plane $\{-\infty<y_1<a,y_0\in\R\}$. 
 Since  the map  (\ref{eq:5.16})   is
the identity  on  $\{r=a,x_0\in\R\}$,  the DN  operators  $\Lambda$ and $\hat\Lambda$  for
(\ref{eq:5.5})  and (\ref{eq:5.18})  are equal  on 
$\{r=a,x_0\in \R\}$,  i.e. 
\begin{equation}                                       \label{eq:5.20}
\Lambda f=\hat \Lambda f\ \ \ \mbox{for all}\ \ f(x_0)\in C_0^\infty(\R).
\end{equation}
Consider  now any metric $g_1$  of the form 
(\ref{eq:5.1})  assuming that conditions  of the form (\ref{eq:4.13}),  (\ref{eq:4.14}),
(\ref{eq:4.15})  are satisfied.
Let
\begin{align}                                            \label{eq:5.21}
&\frac{1}{\sqrt{- g_1(x_0,r)}}\frac{\partial}{\partial x_0}
\Big(\sqrt{-g_1}\, g_1^{00}(x_0,r)\frac{\partial u'(x_0,r)}{\partial x_0}\Big)
+
\frac{1}{\sqrt{-g_1}}\frac{\partial}{\partial x_0}
\Big(\sqrt{-g_1}\,g_1^{0r}\frac{\partial u'}{\partial r}\Big)
\\
\nonumber
+&\frac{1}{\sqrt{-g_1}}
\frac{\partial}{\partial r}\Big(\sqrt{-g_1}\, g_1^{0r}\frac{\partial u'}{\partial x_0}\Big)
+
\frac{1}{\sqrt{-g_1}}\frac{\partial}{\partial r}
\Big(\sqrt{-g_1}\,g_1^{rr}\frac{\partial u'}{\partial r}\Big)
=0,\ r>0, \ x_0\in \R,
\end{align}
Let $r=R_1^+(x_0)$  be the black hole horizon for the metric $g_1$.

Denote  by $D_{g_1}$  the domain  in $[0,a)\times\R^1$  bounded by $r=R_1^+(x_0)$  and  
$\{r=a,x_0\in \R\}$.

Let  $\varphi_1'(x,r),\varphi_2'(x_0,r)$  be  similar  to (\ref{eq:5.9}), (\ref{eq:5.10}),
$$
g_1^{00}\varphi_{1x_0}'(x,r)+b_-'\varphi_{1r}'=0,\ 
g_1^{00}\varphi_{2x_0}'(x_0,r)+b_+'\varphi_{2r}=0,
$$
where  $b_\pm'=g_1^{0r}\pm\sqrt{(g_1^{0r})^2-g_1^{00}g_1^{rr}}$.

Make the changes of variables
\begin{equation}                                              \label{eq:5.22}
y_0'=\frac{\varphi_1'(x_0,r)-\varphi_2'(x_0,r)}{2},\ \ \ 
y_1'=\frac{\varphi_1'+\varphi_2'}{2}
\end{equation}
similar  to (\ref{eq:5.16}), where $\varphi_1',\varphi_2'$
satisfy (\ref{eq:5.15}).  Thus   we transform  the equation (\ref{eq:5.21}) 
 to the equation 
\begin{equation}                                         \label{eq:5.23}
\frac{\partial^2\hat u'(y_0',y_1')}{\partial (y_0')^2}-
\frac{\partial^2\hat u'(y_0',y_1')}{\partial (y_1')^2}
=0,
\end{equation}
in the domain $\{-\infty<y_1'<a,y_0'\in\R\}$.
Note  that  
\begin{equation}                                   \label{eq:5.24}
\hat u'(y_0',y_1')=u'(x_0,r),
\end{equation}
where $(y_0',y_1')$  and  $(x_0,r)$  are related  by (\ref{eq:5.22}).

Assume that DN  operators $\Lambda$  and  $\Lambda'$  
corresponding  to (\ref{eq:5.5})  and  (\ref{eq:5.21})
coincide on $\{r=a,x_0\in\R\}$,  i.e.
\begin{equation}                                     \label{eq:5.25}
\Lambda f=\Lambda' f\ \ \mbox{for all}\ \ f\in C_0^\infty(\R).
\end{equation}
Let $\hat\Lambda'$  be  the DN operator  corresponding  to  (\ref{eq:5.23}).  Since
the map  (\ref{eq:5.22})  is an identity on $\{r=a,x_0\in\R\}$
we have,  as in (\ref{eq:5.20})  that
\begin{equation}                                       \label{eq:5.26}
\Lambda'f=\hat\Lambda' f,\ \ \forall f\in C_0^\infty(\R).
\end{equation}
Therefore   (\ref{eq:5.20}),  (\ref{eq:5.25}),  (\ref{eq:5.26})  imply that 
\begin{equation}                                         \label{eq:5.27}
\hat\Lambda f=\hat\Lambda'f,\ \ \forall f\in C_0^\infty(\R).
\end{equation}
Since  $\hat u'(y_0,y_1)$  and $\hat u(y_0,y_1)$ satisfy  the same equations (\ref{eq:5.18})
and (\ref{eq:5.23}),  respectively, in $\{-\infty<y_1<a,x_0\in\R\}$
 and have  the same Cauchy  data (\ref{eq:5.27}),
we get that
\begin{equation}                                         \label{eq:5.28}
\hat u(y_0,y_1)=\hat u'(y_0,y_1).
\end{equation}
Denote by $(x_0',r')=\sigma(x_0,r)$  
the map of $D_g$  onto  $D_{g'}$  that is a composition  of the map  (\ref{eq:5.16})
and the inverse of the map  (\ref{eq:5.22}).
Then  (\ref{eq:5.18}),  (\ref{eq:5.24})  and  (\ref{eq:5.28})  imply that
\begin{equation}                                 \label{eq:5.29}
u'(\sigma(x_0,r))=u(x_0,r)\ \ \ \ \mbox{in}\ \ D_g.
\end{equation}
The equality  (\ref{eq:5.29})  means  that metrics $g$  and $g'$  
are isometric.  Note that $\sigma(x_0,r)$  sends  the characteristics of (\ref{eq:5.5})
in $D_g$  
to the characteristics of  (\ref{eq:5.21})  in $D_{g'}$.
Thus  the boundary  $r=R^+(x_0)$  of $D_g$  is mapped to  the boundary  
$r=R_1^+(x_0)$of $D_{g'}$ (cf. Theorem \ref{theo:4.2}).
Therefore  the DN  operator of (\ref{eq:5.5})
on $\{r=a,x\in \R\}$  determines  the boundary  $r=R^+(x_0)$ of the black  hole
 up to  an  isometry.

Consider the case  of the white hole,  i.e.  
when  $b_1(x_0)>0$  in  (\ref{eq:4.15}).
Exactly,  as in  the case of black holes,  one can prove that the DN  operator  on  
$\{r=a,x_0\in\R\}$  determines  the boundary  of the white hole  $r=R^-(x_0)$  up
to an isometry.

Let  $D_g^-$ be  the subdomain  of 
$\{0<r<a,x_0\in\R\}$ bounded by  $\{r=a,x_0\in\R\}$  and  the  $r=R^-(x_0)$.
One can  see that  any  solution $u(x_0,r)$ of (\ref{eq:5.2}),
(\ref{eq:5.3}),  (\ref{eq:5.4})  has the support  in $D_g^-$,
i.e.  it is  equal  to zero  when $0<r<R^-(x_0),x_0\in\R$.
Thus boundary measurements on $\partial C=\{r=a,x_0\in\R\}$  can not 
recover any information about the metric in $\{0<r<R^-(x_0),x_0\in\R\}$.

\begin{center}
{\bf Acknowledgment} 
 \end{center}
The author is deeply indebted to the referees for  correcting mistakes and giving many 
valuable suggestions.

\bibliographystyle{amsalpha}

\end{document}